\documentclass[a4paper,twocolumn,fleqn,usenatbib]{mnras}
\usepackage{color}
\usepackage{graphicx}
\usepackage{amssymb,amsmath,epsfig}
\usepackage{natbib}
\input{colordvi.tex}
\definecolor{name}{rgb}{0.16,0.65,0.31}

\usepackage{here}

\begin{document}
\title[Foreground contamination to 21$\,$cm-LAE cross PS]{Detectability of 21$\,$cm-signal during the Epoch of Reionization with 21$\,$cm-Lyman-$\alpha$ emitter cross-correlation. II. Foreground contamination}
\author[S. Yoshiura]{S. Yoshiura$^{1}$,
J.~L.~B. Line$^{2,3}$,
K. Kubota$^{1}$,
K. Hasegawa$^{4}$,
K. Takahashi$^{1}$
\\
\\
$^1$Department of Physics, Kumamoto University, Kumamoto,Japan\\
$^2$The University of Melbourne, Melbourne, Australia \\
$^3$ARC Centre of Excellence for All-sky Astrophysics (CAASTRO) \\
$^4$Department of Physics, Nagoya University, Aichi , Japan\\
}

\date{2017/07/07}

\maketitle
\begin{abstract}
Cross-correlation between the redshifted 21$\,$cm signal and Lyman-$\alpha$ emitters (LAEs) is powerful tool to probe the Epoch of Reionization (EoR). Although the cross-power spectrum (PS) has an advantage of not correlating with foregrounds much brighter than the 21$\,$cm signal, the galactic and extra-galactic foregrounds prevent detection since they contribute to the variance of the cross PS. Therefore, strategies for mitigating foregrounds are required. In this work, we study the impact of foreground avoidance on the measurement of the 21$\,$cm-LAE cross-correlation. We then simulate the 21$\,$cm observation as observed by the Murchison Widefield Array (MWA). The point source foreground is modelled from the GaLactic and Extragalactic All-sky Murchison Widefield Array (GLEAM) survey catalogue, and the diffuse foreground is evaluated using a parametric model. For LAE observations, we assume a large survey of the Subaru Hyper Supreme-Cam (HSC), with spectroscopic observations of the Prime Focus Spectrograph (PFS). To predict the 21$\,$cm signal, we employ a numerical simulation combining post processed radiative transfer and radiation hydrodynamics. Using these models, the signal-to-noise ratio of 2D PS shows the foreground contamination dominates the error of cross-PS even in the so-called `EoR window'. We find that at least 99\% of the point source foreground and 80\% of the galactic diffuse foreground must be removed to measure the EoR signal at large scales $k<0.5\, h\,\rm Mpc^{-1}$. Additionally, a sensitivity 80 times larger than that of the MWA operating with 128 tiles and 99\% of the point source foreground removal are required for a detection at small scales.      
\end{abstract}

\section{Introduction}\label{sec:intro}
Although the epoch of reionization (EoR) is a transformational phase in the history of the universe, it is poorly explored. The birth and growth of the first stars, galaxies, and black holes drove the process of reionization, leaving a footprint in the EoR signal. One of the most active challenges of cosmology today is to reveal the evolution of the intergalactic medium during EoR. 

There are broad observational constraints on the process of reionization in the literature. For example, the spectra of high redshift quasars have proved that the EoR was completed by $z\sim6$~\citep{2006AJ....132..117F, 2015MNRAS.447..499M}. The analysis of the cosmic microwave background has found that the redshift of the half-point evolution of the EoR was around $7.8<z<8.8$~\citep{2016A&A...596A.108P}. The evolution of luminosity functions of high redshift Lyman-$\alpha$ emitters (LAEs) at $z>6$ has constrained the neutral fraction of hydrogen~\citep{2010ApJ...723..869O,2017arXiv170407455O, 2014ApJ...797...16K, 2017arXiv170501222K,2017arXiv170408140S}. 

During and before the EoR, the IGM is rich in neutral hydrogen (\ion{H}{i}). Thus, for revealing a more precise evolution of the neutral fraction, understanding the nature of the main sources of reionization, the redshifted 21$\,$cm signal emitted from \ion{H}{i} gas is powerful tool. The redshifted 21$\,$cm signal allows the creation of three dimensional images of 21$\,$cm signal, reflecting the \ion{H}{i} gas density and its neutral fraction. There are various previous works which studied the usefulness of, for example, the topology of 21$\,$cm signal \citep{2006MNRAS.370.1329G, 2011MNRAS.413.1353F, 2017MNRAS.465..394Y, 2017MNRAS.471.1936K}, one point statistics \citep{2014MNRAS.443.3090W, 2015MNRAS.454.1416W, 2016PASJ...68...61K}, power spectrum \citep{2004ApJ...613....1F,2006ApJ...653..815M,2007MNRAS.376.1680P, 2014ApJ...782...66P, 2015MNRAS.451..467S}, and bispectrum \citep{2015MNRAS.451..266Y, 2016MNRAS.458.3003S, 2017MNRAS.468.1542S, 2017arXiv170808458M}. However, the 21$\,$cm signal has yet to be directly observed.

Measuring the 21$\,$cm signal is promising and therefore there are a number of ongoing experiments such as the the Giant Metrewave Radio Telescope EoR Experiment (GMRT, \cite{2013MNRAS.433..639P}), the Donald C. Backer Precision Array for Probing the Epoch of Reionization (PAPER, \cite{2010AJ....139.1468P}), the LOw Frequency ARray (LOFAR, \cite{2013A&A...556A...2V}), and the Murchison Widefield Array (MWA, \cite{2013PASA...30....7T,2013PASA...30...31B}). Meanwhile, future experiments to observe the 21$\,$cm signal such as the Hydrogen Epoch of Reionization Array (HERA, \cite{2016icea.confE...2D}) and the low frequency Square Kilometre Array (SKA\_{}LOW, \cite{2013ExA....36..235M}) are due to be constructed in few years. Because the sensitivities of the current generation of telescopes are not sufficient for imaging the 21$\,$cm intensity map, these experiments target a detection of the the power spectrum of 21$\,$cm signal (here after PS stands for power spectrum). Upper limits on the 21$\,$cm PS have been provided~\citep[e.g.][]{2015ApJ...809...61A, 2015PhRvD..91l3011D, 2014ApJ...788..106P, Jacobs2015,2016ApJ...833..102B,2017ApJ...838...65P}. These limits are either consistent with the expected sensitivity of the telescope, or limited by imperfect calibration and modelling of astrophysical foregrounds. Measuring the 21$\,$cm signal is extremely challenging due the existence of bright foregrounds due to the synchrotron radiation of our own galaxy and extra galactic sources, which are three orders of magnitude brighter than the expected 21$\,$cm signal. Thus, one needs to precisely remove the foregrounds, but the process is complicated by instrumental systematics and calibration errors, which prevent the detection of the 21$\,$cm signal. 

There are three main strategies to separate the foregrounds from the 21$\,$cm signal, which we summarise here: 
\begin{enumerate}
\item \textit{Direct foreground removal} - By fitting the foregrounds as a function of frequency, either by fitting a polynomial or using non-parametric methods, one can subtract the fitted foregrounds from the data. These methods have been used on real data and have had moderate levels of success but the detection has not yet been achieved.
\item \textit{Foreground avoidance} - The foregrounds should be spectrally smooth with frequency, and therefore should be sub-dominant in Fourier space to a signal that is not spectrally smooth, such as the expected EoR signal, at high Fourier modes. The Fourier-transform w.r.t frequency is labelled $k_{\parallel}$, as the frequency response probes spatial scales parallel to the line of sight, via the redshifting of the 21$\,$cm signal. The foregrounds are therefore expected to be confined to low $k_{\parallel}$, leaving an EoR `window' in which the 21$\,$cm signal and the foreground contamination are on order of the same power. However, mode-mixing due to the chromaticity of an interferometer causes the foregrounds to re-distribute into high $k$-modes, into an area of $k$-space called the `wedge'~\citep{2010ApJ...724..526D,2012ApJ...752..137M,2012ApJ...757..101T}.
\item \textit{Cross-coleelation with other measurements} - In this work, we study the cross-correlation between the 21$\,$cm signal and LAEs. As LAEs are one of the main ionizing sources during the EoR, the distribution of LAEs correlates with 21$\,$cm signal. LAEs also have the advantage of not strongly correlating with foreground sources. There are previous works which investigate the cross-correlation and its feasibility~\citep{2007ApJ...660.1030F, 2009ApJ...690..252L,2013MNRAS.432.2615W,2014MNRAS.438.2474P,2016MNRAS.459.2741S,2016arXiv161109682H,2017ApJ...836..176H,2017arXiv170107005F,Kubota}. 
\end{enumerate}

In previous studies of the 21$\,$cm - LAE correlation, foreground sources have been ignored since the foregrounds should have no correlation with the LAE distribution. {Then, the cross PS of foregrounds and LAE is zero. None the less, its non-zero statistical fluctuation, that is, its variance  contributes to the error to the 21$\,$cm - LAE cross PS. }
Moreover, the power of the foregrounds are large compared to the 21$\,$cm - LAE correlation signal, and therefore possibly the strongest error for the 21$\,$cm - LAE cross PS. Thus, for designing future observational strategies, it is informative to quantitatively estimate the foreground contamination using realistic simulations of the 21$\,$cm signal, the foregrounds, and instrumental effects.

This is the second article in our series on the feasibility of studying the cross correlation between the 21$\,$cm signal and LAEs. In our previous article \cite{Kubota}, we have shown the 21$\,$cm-LAE cross PS can be observed at large scales with either MWA or SKA\_{}LOW 21$\,$cm observations, in tandem with Subaru Hypersuprime Cam (HSC)-Prime Focus Spectrograph (PFS) LAE observations. In addition, the combination of SKA\_{}LOW and HSC-PFS LAE observation has potential to measure the cross PS at small scales, as well as the turn over scale which is identified as a sign transition of the cross PS. Recently, a HSC project dubbed the Systematic Identification of LAEs for Visible Exploration and Reionization Research Using Subaru HSC (SILVERRUSH) found thousands of LAEs at high-$z$. The PFS is a spectrograph system mounted at the prime camera of the Subaru telescope which is due to start observations in 2020. The PFS can precisely identify the redshifts of these LAEs. 

However, in \cite{Kubota}, foregrounds have not been taken into account. In this work, we investigate the impact of foregrounds on the measurement of the 21$\,$cm-LAE cross PS. We simulate an observation of the 21$\,$cm signal with the MWA, use the LAE survey with the HSC, and assume the future spectroscopic measurement of LAEs with PFS. The foreground model is built from a catalogue of real radio point sources, and a physically motivated diffuse emission model. We also employ a contemporary reionization model which combines large scale cosmological radiative transfer with precise radiative hydrodynamical effects on small scales. Finally, we reveal the feasibility of a first detection of 21$\,$cm signal from the EoR, by creating the cross PS, focussing on the EoR window and foreground removal. 

This paper is structured as follows. In Sec.~\ref{sec:method}, we outline our method for calculating the 21$\,$cm-LAE cross PS and error. In Secs.~\ref{sec:MWA} and~\ref{sec:FGmodel}, we introduce the MWA and our foreground model. In Sec.~\ref{sec:signal}, we describe our reionization and LAE models. In Sec.~\ref{sec:result}, we show the 2D PS, the 1D PS and signal to noise ratio. Finally, we summarise our work in Sec.~\ref{sec:summary}. We assume a $\Lambda$CDM cosmology with parameters ($\Omega_{\rm m}$, $\Omega_{\rm \Lambda}$, $H_0$, $h$) = (0.3, 0.7, 68$\,\rm km\,s^{-1}\,Mpc^{-1}$, 0.68). 

\section{Methodology}\label{sec:method}
In this section, we introduce the 21$\,$cm-LAE cross PS and detail the error formalism. Our error formalism is identical to that of \cite{Kubota} except for foreground contamination; for more details on the formalism see \cite{Kubota}. The observed 21$\,$cm brightness temperature $\delta T_{\rm b}$ is written as \citep[e.g.][]{2006PhR...433..181F}):
\begin{eqnarray}
\delta T_{\rm b}(z)
\approx 27 x_{\rm \ion{H}{i}} (1 + \delta_{\rm m})
          \bigg(1 - \frac{T_{\gamma}}{T_{\rm S}}\bigg)\nonumber\\
          \times\bigg(\frac{1+z}{10} \frac{0.15}{\Omega_{\rm m} h^2}\bigg)^{1/2}
          \bigg(\frac{\Omega_{\rm b} h^2}{0.023} \bigg) ~ [{\rm mK}],
\label{eq:brightness}
\end{eqnarray}
where $x_{\rm \ion{H}{i}}$ is the neutral fraction of \ion{H}{i}, $\delta_{\rm m}$ the matter over-density, $T_{\rm \gamma}$ the CMB temperature, and $T_{\rm S}$ the spin temperature. 

The cross PS of the 21$\,$cm brightness temperature and the number density of LAEs is defined as
\begin{eqnarray}
\langle \tilde{\delta T_{\rm b}} (\bold{k}) \tilde{\delta}_{\rm gal}(\bold{k}')\rangle = (2\pi)^3 \delta_{\rm D}(\bold{k}+\bold{k}') P_{\rm 21,gal}(\bold{k}),
\end{eqnarray}
where tildes represent Fourier space quantities, $\delta_{\rm D}$ is the Dirac delta function and $\langle\rangle$ denotes an ensemble average. The galaxy over-density is calculated as $\delta_{\rm gal}({\bold{r}}) = n_{\rm gal}({\bold{r}})/\bar{n}_{\rm gal}-1$, where $n_{\rm gal}({\bold{r}})$ is the number density of LAEs. The 1D PS is defined as $\Delta_{\rm 21,gal}(k) = {k^3}P_{\rm 21,gal}(k)/{2\pi^2}$.

The error formula of the cross PS measurement was built in previous works: \citet{2007ApJ...660.1030F, 2009ApJ...690..252L,2014MNRAS.438.2474P,Kubota}. However, the contribution from the foreground PS has always been ignored. In this work, we add a foreground term, meaning the error can be estimated as
\begin{eqnarray}
2\sigma^2_{\rm 21,gal} =  P^2_{\rm 21,gal} + (P_{\rm 21}+N_{\rm 21}+P_{\rm FG})(P_{\rm gal}+N_{\rm gal}).
\label{crossnoise}
\end{eqnarray}
The $P_{i}$ represents the PS of the 21$\,$cm signal, foregrounds, LAEs (denoted as 21, FG, gal respectively). 
It should be noted that the foreground PS is a sum of the contribution from Galactic diffuse emission and extra galactic point sources. The thermal noise of the telescope used to measure the 21$\,$cm signal, $N_{\rm 21}$, is explained in Sec. ~\ref{sec:21cmnoise}. $N_{\rm gal}$ is the shot noise on the observations of the LAEs, which is written as 
\begin{eqnarray}
N_{\rm gal}=\bar{n}_{\rm gal}^{-1}\exp\left(k_{\parallel}^2\frac{c^2\,\sigma^2_{z}}{H^2(z)}\right),
\end{eqnarray}
where $\bar{n}_{\rm gal}$ is the averaged number density of LAEs and $\sigma_{z}$ is the redshift error. {In this work, we assume the PFS observation which has $\sigma_{z}=0.0007$ \citep{2014PASJ...66R...1T,2016SPIE.9908E..1MT}.} For comparison in the following sections, it is worth noting that the thermal noise and shot noise in Eq.~\ref{crossnoise} are detection limiting errors, as these errors are unavoidable. \cite{Kubota} found that thermal noise and the LAE PS dominate the error of cross PS ignoring the foreground term.

{
The 21$\,$cm signal is observed using a radio interferometer, the instrumental outputs of which can be split into frequency and angular direction on the sky. By observing the 21$\,$cm signal at different frequencies (and therefore different redshifts), interferometers measure the 21$\,$cm signal along the line of sight. The 21$\,$cm signal should not be smooth with frequency, due to the distribution of \ion{H}{i} along the line of sight. On the other hand, the foregrounds should be spectrally smooth, and so the foreground PS is expected to be weak at high $k$ modes in the spectral dimension. Thus, in following sections, we use the 2D PS. The 2D PS splits the $k$-mode into those derived from the angular response which is perpendicular to the line of sight $k_{\perp}$, and derived from the frequency which is parallel to the line of sight $k_{\parallel}$. Here, the wave number of line of sight is described as $k_{\parallel}=\mu k$, where $\mu$ is cosine of angle between line of sight and $\bold{k}$. } In the 2D plane, redshift space distortions enhance the PS as $P(k_{\perp},k_{\parallel}) = (1+\beta \mu^2)^2 P(k)$ \citep{1987MNRAS.227....1K}. Here, $\beta$ is $\Omega_m^{0.6}/b$, where $b$ is the bias. {We assume {$b(k)=\sqrt{P_{\rm gal}(k)/P_{\rm DM}(k)}$} for the PS of LAEs, {where $P_{\rm gal}(k)$ and $P_{\rm DM}(k)$ are calculated using simulation data}. Following previous works, we set $b=1$ for the 21$\,$cm signal.}

The error on the cross PS decreases when it is averaged over individual $k$-modes. The reduced error is estimated as 
\begin{eqnarray}
\frac{1}{\sigma^2_{\rm 21,gal}(k)}=\sum_{\mu} \frac{\Delta k \Delta \mu k^2V_{\rm sur}}{4\pi^2} \frac{1}{\sigma^2_{\rm 21,gal}(k,\mu)},
\label{2D21D}
\end{eqnarray}
where $V_{\rm sur}=S_{\rm sur}\Delta D$ is survey volume, $S_{\rm sur}$ is survey area and $\Delta D$ is the survey depth. The deep survey of the HSC covers an area of $27\rm\,deg^2$, which is much smaller than the MWA survey area of $800\rm\,deg^2$. We therefore assume $S_{\rm sur} = 27\rm\,deg^2$ of HSC as a fiducial value in this work. The survey depth is proportional to bandwidth. 

Interferometers naturally output visibility data $V(u,v,w,\nu)$ which is obtained as a function of the separation between receiving elements in the interferometer, $u,v,w$, and frequency, $\nu$. {The mode mixing effect on the FG PS is crucial to understanding foreground contamination, and the thermal noise PS depends on the array configuration of the interferomenter. It is therefore best to simulate these effects in the natural observational space of the intereferometer, rather than directly in PS space.} In the following sections we describe the thermal noise and point source and diffuse foregrounds, all of which are calculated from visibilities. In rest of this section, we introduce the PS calculated from visibilities. 

We prepare a 3D grid of $(u,v,\nu)$ and interpolate all visibilities to this grid. Here, we ignore the $w$-term for simplicity. To calculate the 2D PS, the gridded data are first Fourier transformed along frequency. We define this Fourier transformation as 
\begin{eqnarray}
\tilde{V}(u,v,\eta) = \int V(u,v,\nu) W_B(\nu)\exp(-2\pi i \nu\eta)d\nu,
\end{eqnarray}
where $W_B$ is the bandpass window function, {which acts to surpress foreground leakage from low $k$-modes into higher $k$-modes}. We employ the Blackman Harris Window function as $W_B$ to effectively reduce foreground leakage into EoR window \citep{2013ApJ...776....6T, 2016ApJ...825....9T}. {We note that the amplitude of the foregrounds in the EoR window depend on the window function. For example, foreground contamination increases by orders of magnitude in the EoR window with a rectangular window function. Thus, the error of 21$\,$cm - LAE cross PS increases without the Blackman Harris Window function.} 

The PS is calculated from visibilities as
\begin{eqnarray}
P(k_{\perp},k_{\parallel}) = |\tilde{V}(u,v,\eta)|^2\left(\frac{A_e}{\lambda^2\Delta B}\right)\left(\frac{D^2_{m}\Delta D}{\Delta B}\right),
\label{eq:PSvis}
\end{eqnarray}
where $\eta$ is the Fourier dual of frequency, $\lambda$ is the observed wavelength, $A_e$ is the effective area of antennae, $\Delta B$ is the bandwidth, $D_m$ is the transverse comoving distance and $\Delta D$ is the comoving width corresponding to the bandwidth.

{We note we use a bandwidth of 8$\,$MHz in Equation.~\ref{eq:PSvis} for the MWA. The filter width of the HSC however is $\sim2.5\,$MHz, which corresponds to 40$\,$cMpc at $z\sim6.6$, which we use in Equation.~\ref{2D21D}}.

\section{MWA}\label{sec:MWA}\label{sec:21cmnoise}
The Murchison Widefield Array (MWA) is an interferometric array, consisting of 128 `tiles' over a radius of $\sim1500\,$m. Each tile is a $4\times4$ arrangement of dipole antennae, with an observable frequency range 80 - 300$\,$MHz. The output data resolutions are 40$\,$kHz in frequency and 0.5$\,$s in time. One of the MWA primary science goals is the measurement of the EoR 21$\,$cm signal. In this work, we assume that the MWA observes a foreground quiet region, labelled the EoR0 field at RA $0^h$, Dec $-27^\circ$. We assume a 3 hour continuous observation around zenith and calculate visibilities every 8$\,$s. The time interval is 16 times larger than the actual MWA resolution; we choose this interval to reduce our computational costs.

\begin{figure}
\centering
\includegraphics[width=8.5cm]{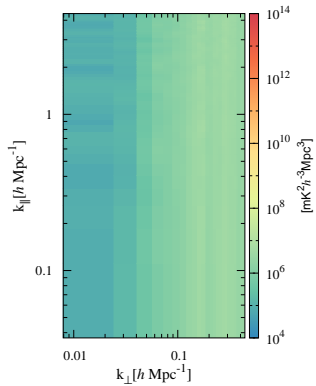}
\caption{{The expected thermal noise calculated from visibilities as described in Sec.~\ref{sec:MWA}. The MWA $uv$-coverage becomes sparse beyond the core and the noise increases at high $k_{\perp}$. }} 
\label{fig:Noise} 
\end{figure}

As shown in \cite{Kubota}, the thermal noise of 21$\,$cm experiments is one of the dominant sources of error on the cross PS. The thermal noise for a visibility can be written as
\begin{eqnarray}
\sigma_{21} = \frac{\lambda^2}{A_{\rm eff}}\frac{T_{\rm sys}}{\sqrt{\Delta t \Delta \nu}},
\end{eqnarray}
{where the $\sigma_{21}$ consists of the system temperature $T_{\rm sys}$, integration time per visibility $\Delta t$, and channel width $\Delta \nu$. For the MWA, the effective antenna area $A_e$ is 14$\rm\,m^2$.} 
{To increase our integration time, we assume we make the same 3$\,$hour observation 333 times, effectively creating a $\sim999$ hour observation. As we originally simulated our visibilities with a time resolution of 8$\,$s, this sets $\Delta t = 8 \times 333 = 2664\,$s.} For each simulated visibility, we add thermal noise by randomly drawing a value from a gaussian distribution with standard deviation $\sigma_{21}$, {where ($\Delta \nu$ , $T_{\rm sys}$, {$\Delta t$}) = (80$\,$kHz, 289$\,$K, 2664$\,$s).} Fig.~\ref{fig:Noise} shows the expected thermal noise of the MWA. As the added thermal noise is uncorrelated between visibilities, the thermal noise contribution decreases with increased $uv$ sampling. This is shown in Fig.~\ref{fig:Noise}, as the noise decreases at small $k_{\perp}$, which reflects the large  $uv$-coverage of MWA at small $k_{\perp}$ due to its numerous short baselines.


\section{Foreground models}\label{sec:FGmodel}
Although there are various components that act as foregrounds to the 21$\,$cm signal, such as synchrotron emission and free-free emission, here we take into account contributions from extra galactic point sources and Galactic synchrotron emission, which are expected to be the dominant component of the foregrounds. In this section a realistic point source model and a reasonable diffuse emission model are described. 

\subsection{Point sources}\label{sec:point}
We base our point source model on the GLEAM catalogue~\citep{2017MNRAS.464.1146H}, which covers most of the sky south of declination 30$^\circ$, excluding difficult survey areas such as the Galactic plane and the Magellanic clouds. The GLEAM survey~\citep{Wayth2015} was undertaken using the MWA and so makes a natural choice for this work. The catalogue contains 307,455 sources, of which 245,470 sources are reported with a fitted spectral index (SI). The SI, $\alpha$, relates the flux density of a source, $S$, to the frequency through $S\propto \nu^\alpha$. To assign realistic SIs to the remaining 61,985 sources, we fit a normal distribution to the existing SI values, and then draw random values from this fitted distribution. We fit a normal distribution with $\mu = -0.81, \sigma = 0.24$. For simplicity, we assume all point sources follow this simple power law, however in reality a significant fraction of sources have more complicated spectral behavior such as gigahertz-peaked spectrum (GPS) and compact-steep spectrum (CSS) sources~\citep[for further details see][and references within]{Callingham2017}. With the positional and spectral information, we are able to estimate the flux density across most of the sky, at all frequencies.
We generate a 3$\,$hour observation's worth of visibilities using OSKAR\footnote{\url{http://oskar.oerc.ox.ac.uk/}}~\citep{Mort2010}, which is a GPU-enabled interferometric simulation package. We run our mock observation with the EoR0 field centre initially at an hour angle of $-1.5^h$, and set the MWA to observe in 2$\,$minute snapshot pointings over the 3$\,$hour observational period.

\begin{figure}
\centering
\includegraphics[width=8.5cm]{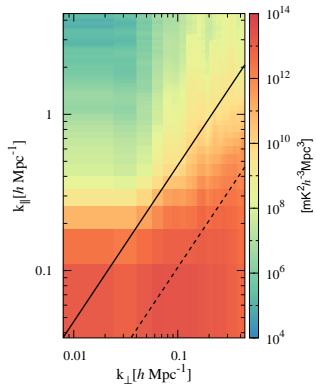}
\caption{{A 2D PS of the simulated point sources. {The solid and dashed lines show the expected foreground contamination limits caused by point sources at the horizon and the edge of the primary beam, respectively.} The figure clearly shows the wedge feature and the power is reduced effectively outside the horizon limit. The leakage of power at $k_{\perp}\sim0.15$ is caused by insufficient $uv$-sampling.}} 
\label{fig:Jpfg} 
\end{figure}

Fig.~\ref{fig:Jpfg} shows the 2D PS of point sources, which clearly shows the foreground wedge and the EoR window structure. We can find leakage of foreground power into the EoR window, which is caused by the discrete sampling of $uv$-data along frequency. In particular, the insufficient number of long baselines shifts power from within the wedge into the window at $k_{\perp}\sim0.15\,h\,\rm Mpc^{-1}$. The two diagnostic lines plotted in Fig.~\ref{fig:Jpfg} represent the expected foreground contamination limits caused by point sources at the observational horizon, and the edge of the MWA primary beam (solid and dashed lines respectively).

\subsection{Diffuse emission}\label{sec:diffuse}
\begin{figure}
\centering
\includegraphics[width=8.5cm]{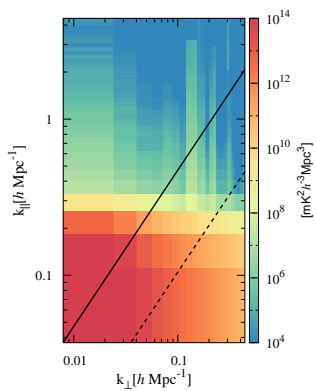}
\caption{A 2D PS of the diffuse foreground model. The diffuse emission has strong contamination beyond horizon limit at $k_{\parallel} = 0.1, k_{\perp}=0.01$. The vertical streaks are due to sparse $uv$-coverage of the MWA at high $k_{\perp}$.}
\label{fig:Cdfg} 
\end{figure}

Galactic magnetic field lines and free electrons in the interstellar medium interact and emit synchrotron radiation. The emission is expected to have a smooth spectral response and power on large spatial scales. We employ the model and parameters found in \cite{2008MNRAS.389.1319J}. The PS can be written as~\citep[c.f.][]{2016ApJ...818..139T} 
\begin{eqnarray}
P_{\rm FG,D} = (\eta T_{\rm FG,D})^2 \left(\frac{\rm u}{\rm u_0}\right)^{-2.7}\left(\frac{\nu}{\nu_0}\right)^{-2.55},
\end{eqnarray}
where $T_{\rm FG,D}=235\rm\,K$ is the average temperature of the diffuse emission with the fluctuation fraction, $\eta=0.01$. The PS follows a power law in angular scale, ${\rm u}=(u^2+v^2)^{1/2}$, and frequency, $\nu$, with $\rm u_0=10\,\lambda$ and $\nu_0=100\rm MHz$. We mention that the diffuse foregrounds are symmetric about $(u,\,v)=(0,\,0)$.
Fig.~\ref{fig:Cdfg} shows the PS of diffuse foreground. Although the diffuse emission is larger than $10^{15}$ at large scale, the power deceases to $10^{6}$ in the EoR window due to the smooth spectra. The vertical lines at $k\sim0.15\,h\,\rm Mpc^{-1}$ are caused by missing $uv$-samples, a natural consequence of the $uv$-coverage of the MWA baseline distribution. {The same structure is faintly shown in Fig.~\ref{fig:Jpfg}. }

\section{Signal model}\label{sec:signal}
In this section, we describe the reionization simulation used in this work, and how we choose the LAEs including the effect of Lyman-$\alpha$ transmission. The IGM data used is identical to those in \cite{Kubota} and more details are found there \citep[see also][]{2016arXiv160301961H}.

Radiation hydrodynamical (RHD) simulations have found that radiative feedback processes affect the star formation rate within galaxies, and also affect the clumping factor of gas in the IGM.  Although these effects should be taken into account in large-scale cosmological simulations, performing RHD simulations within volumes larger than a $\rm 100\,Mpc$ cubed box is difficult due to the prohibitive computational costs. Therefore, we model the results of the RHD simulation and adapt it into a large volume post processing radiative transfer (RT) simulation.

The RHD simulation is performed using $2\times512^3$ particles in a 20$\rm\,Mpc$ cubed box. The simulation shows that the escape fraction of galaxies is affected by UV photons and supernovae feedback, and that lower mass galaxies have higher escape fractions. The simulation also shows that the clumping factor depends on the local density as well as the local ionized fraction. {Here, the escape fraction controls the ionizing efficiency of each galaxies and this directly relates to the distribution of ionized regions. Also, the inhomogeneous clumping factor determines the recombination rate of ionized IGM and provides inhomogeneous 21$\,$cm signal distribution. These two factors are important to not only the 21$\,$cm signal distribution but also the 21$\,$cm-LAE cross PS.} Therefore, in order to include these result in the RT simulation, we make a look up table of the spectral energy distribution of galaxies and the clumping factor depending on the halo mass, ionized fraction and local density.

In order to obtain the matter density and halo distribution, an N-body simulation is performed with a massive parallel TreePM code, GreeM \citep{2009PASJ...61.1319I,2012arXiv1211.4406I}, with $4096^3$ particles inside a $\rm160\,Mpc$ cubic box. For the RT simulation, we separate the volume up into $256^3$ uniform grids. Using the gridded result, we solve the ionization equation of \ion{H}{i}, neutral and ionized Helium (\ion{He}{i} and \ion{He}{ii}), and the thermal equation, simultaneously. The spectral energy distribution of galaxies and recombination rates are evaluated by referring to the results of the RHD simulation. The differential brightness temperature distribution is estimated from the neutral fraction and matter density distribution. We then use Eq.~\ref{eq:brightness} and assume that the spin temperature is completely coupled with the gas temperature.

In this work, we employ `mid' and `late' models for estimating the cross PS. In the late model, we reduce the ionizing efficiency to be 1.5 times lower than that of mid model. These models satisfy the constraints on the ionized fraction at $z\sim6$ indicated from quasar spectra and the CMB optical depth due to Thomson scattering \citep[]{2006AJ....132..117F, 2015MNRAS.447..499M,2016A&A...596A.108P}. The mid model has the averaged brightness temperature $\bar{\delta T_b}=0.21\,\rm mK$ and the volume averaged neutral fraction $\bar{x}_{\rm \ion{H}{i}}=0.017$. The late model has $\bar{\delta T_b}=6.7\,\rm mK$ and $\bar{x}_{\rm \ion{H}{i}}=0.44$. For more details of these models, please see \cite{Kubota}. 

Finally, we detail how we define observed LAEs from our simulation. Based on the results of the RHD simulation, we estimate the intrinsic Lyman-$\alpha$ luminosity $L_{\rm \alpha,int}$ of each galaxy. We find that $L_{\rm \alpha,int}$ of galaxies more massive than $10^{10}\,M_\odot$ follow below relation
\begin{eqnarray}
L_{\rm \alpha,int}=10^{42}(M_{h}/10^{10})^{1.1},
\end{eqnarray}
where $M_{h}$ is the halo mass of the galaxy. However, when deriving this relation, we have ignored absorption of Lyman-$\alpha$ photons by dust in the ISM. We therefore take it into account by introducing the escape fraction of Lyman-$\alpha$ photons from a galaxy as a free parameter, $f_{\alpha}$. 

As a next step, we estimate a transmission rate of Lyman-$\alpha$ photons, $T_{\alpha}$. The intrinsic line profile of Lyman-$\alpha$ photons (depending only on the nature of galaxy) is obtained from a Lyman-$\alpha$ RT calculation with an expanding spherical cloud model. The radial velocity of gas is evaluated as $v(r)=V_{\rm  out}(r/r_{\rm vir})$, where $V_{\rm out}$ and $r_{\rm vir}$ are the galactic wind velocity and the virial radius of a halo. The shape of the intrinsic line profile depends on $V_{\rm out}$ and the \ion{H}{i} column density in a galaxy~\citep[see][for details]{2017arXiv170105571Y}. Based on the line profile, $T_{\alpha}$ is evaluated by integrating over an optical depth derived from a line of sight through 80$\,$cMpc of the IGM from a galaxy.

Using this prescription, the observed Lyman-$\alpha$ luminosity is evaluated as
\begin{eqnarray}
L_{\rm \alpha,obs}=f_{\alpha} T_{\alpha} L_{\rm \alpha,int}. 
\end{eqnarray}
In this work, we set parameters $f_{\alpha}$, $V_{\rm out}$, and $N_{\rm \ion{H}{i}}$ so that the simulated Lyman-$\alpha$ luminosity function corresponds to observations in \citet{2017arXiv170501222K}. The parameter set is identical to that in \cite{Kubota} and a comparison of the luminosity functions is shown in their Fig. 2.

\section{Results}\label{sec:result}
Using the models described in previous sections, we now can calculate the cross PS and the signal to noise ratio. We also discuss the requirement for detecting the cross PS.

\subsection{2D Power Spectrum}\label{sec:2Dpk}
\begin{figure*}
\centering
\includegraphics[width=8.5cm]{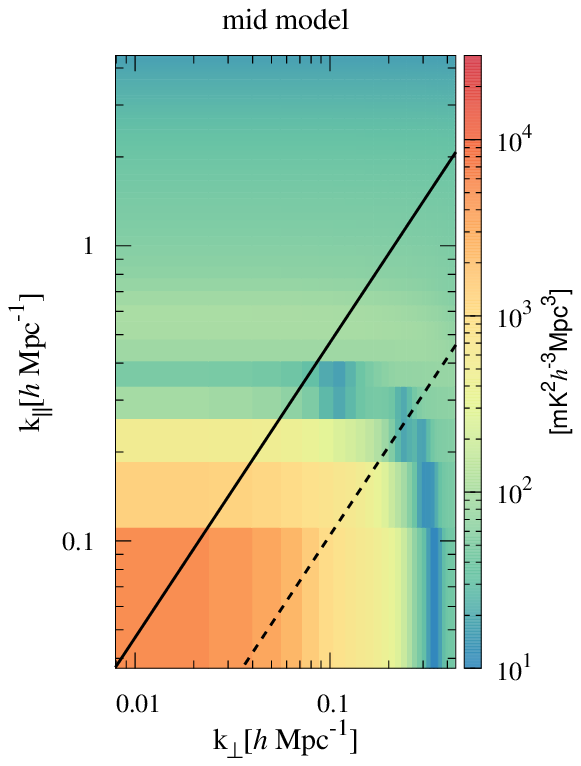}
\includegraphics[width=8.5cm]{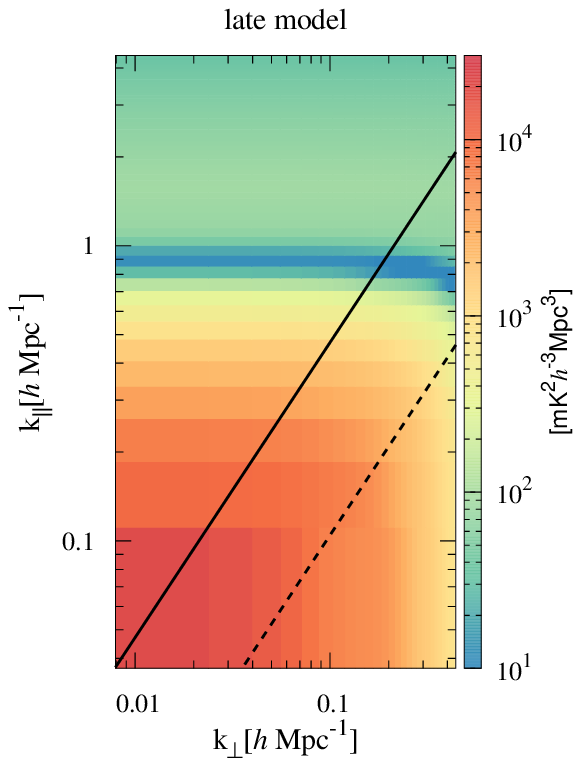}
\caption{{A 21$\,$cm-LAE cross PS computed from our simulation. The mid model is plotted to the left, the late to the right. Blue curves indicate a sign transition in the correlation, called the turn over scale. The turn over scale is $0.3{h\,\rm Mpc^{-1}}$ for the mid model and $0.9{h\,\rm Mpc^{-1}}$ for the late model. The amplitude shown is the absolute value; the actual signal has a negative value at low $k$-modes (large spatial scales).}} 
\label{fig:2Dpk} 
\end{figure*}

The left and right panels of Fig.~\ref{fig:2Dpk} show the 2D cross PS of mid and late models, respectively. The signal of the mid model is $\sim$ one order of magnitude smaller than that of the late model. The difference can be understood as the difference of mean values of the brightness temperature. The cross PS tends to be powerful at small $k$ for both models.
 
The cross PS has negative values at small $k$ since large ionized bubbles are created around LAEs, and these cause a negative correlation at large scales. For the late model, the horizontal stripe seen at $k_{\parallel}=0.9\,h\,\rm Mpc^{-1}$ indicates a sign transition from negative to positive, called the turn over scale, which represents the typical size of ionized bubbles. The turn over scale of the mid model is $k\sim0.3\,h\,\rm Mpc^{-1}$. This indicates that the typical bubble size of mid model is larger than the size of late model. 

In these figures, we show the absolute value of the cross PS using a logarithmic scale. Conversely to the large scale correlation, the cross PS has a positive correlation at small scales. The positive value can be explained as the correlation between clumps of \ion{H}{i} near haloes, with the overall matter density, and the LAE distribution. 

\begin{figure*}
\centering
\includegraphics[width=8.5cm]{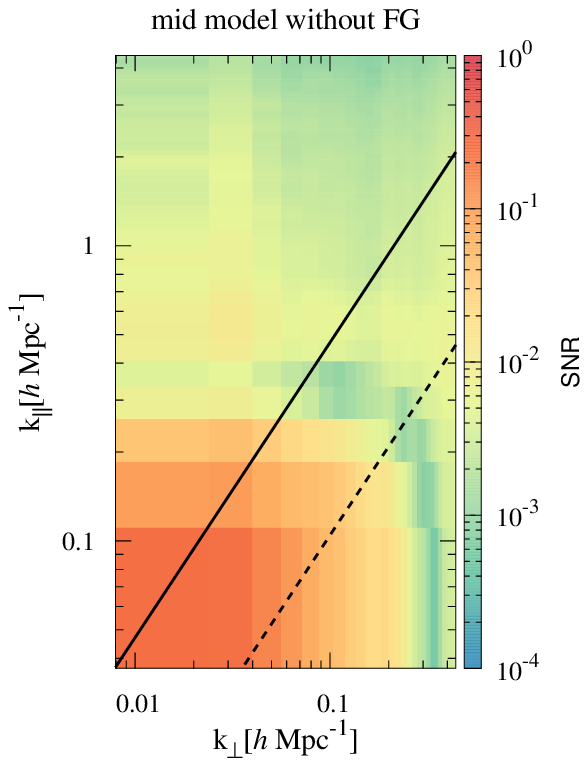}
\includegraphics[width=8.5cm]{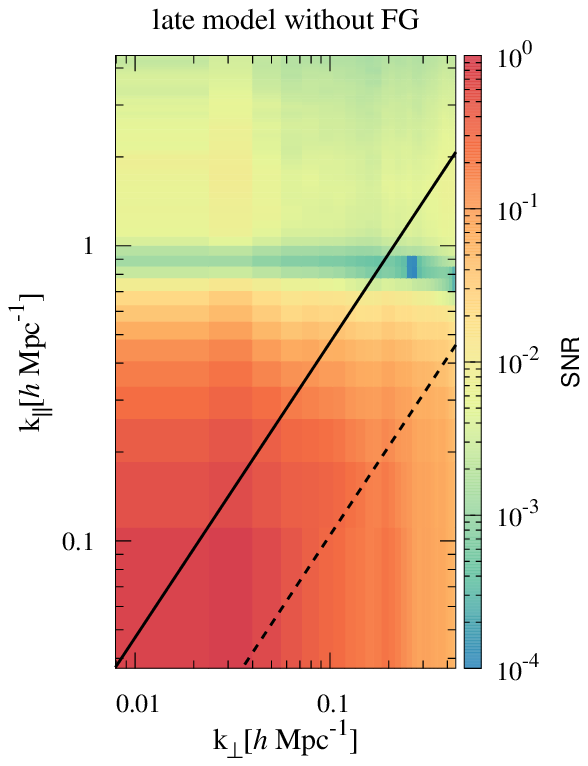}
\caption{{The SNR without foreground contamination for the mid and late models. The structure of SNR resembles the signal except at $k_{\perp}<0.03\,h\, \rm Mpc^{-1}$ where the thermal noise decreases with the increase of the baseline number density of the MWA.}} 
\label{fig:nofgSNR} 
\end{figure*}

Visualizing the signal to noise ratio (SNR) is useful to explore at which scales we can detect the signal. Here, the error on the cross PS is estimated using Eq.~\ref{crossnoise}. The plots in Fig.~\ref{fig:nofgSNR} represent the SNR without contamination of foregrounds. As shown in \cite{Kubota}, the thermal noise and galaxy PS are the dominant sources of error, and have smooth structure. Thus, the structure of SNR resembles the signal, except where $k_{\perp} \lesssim 0.03\, h\,\rm Mpc^{-1}$, where the SNR increases because the thermal noise decreases due to the higher baseline density of MWA at small $k_{\perp}$. For the late model, the SNR is $>1$ at $k_{\perp}<0.03\,h\,\rm Mpc^{-1}$ and $k_{\parallel}<0.1\,h\,\rm Mpc^{-1}$, which indicates the MWA can detect the signal with perfect foreground removal. However, this scale is in the foreground wedge where the galactic synchrotron emission is extremely large compared to the signal.

\begin{figure*}
\centering
\includegraphics[width=8.5cm]{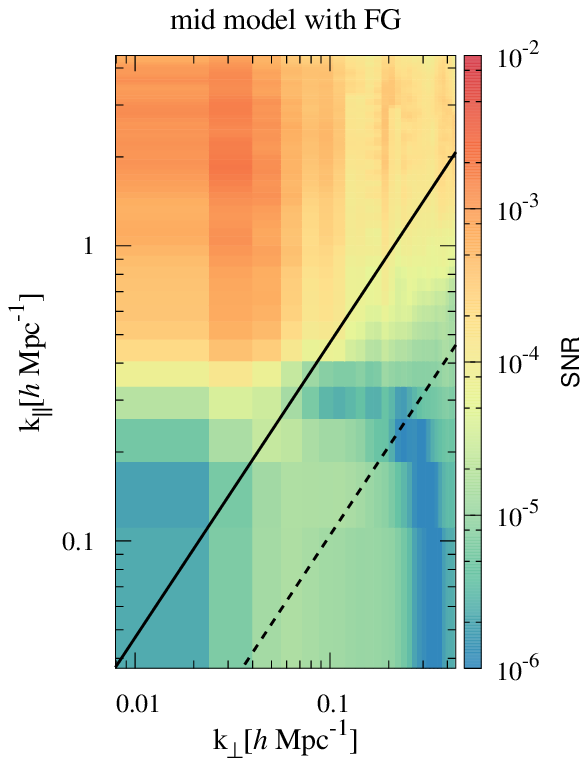}
\includegraphics[width=8.5cm]{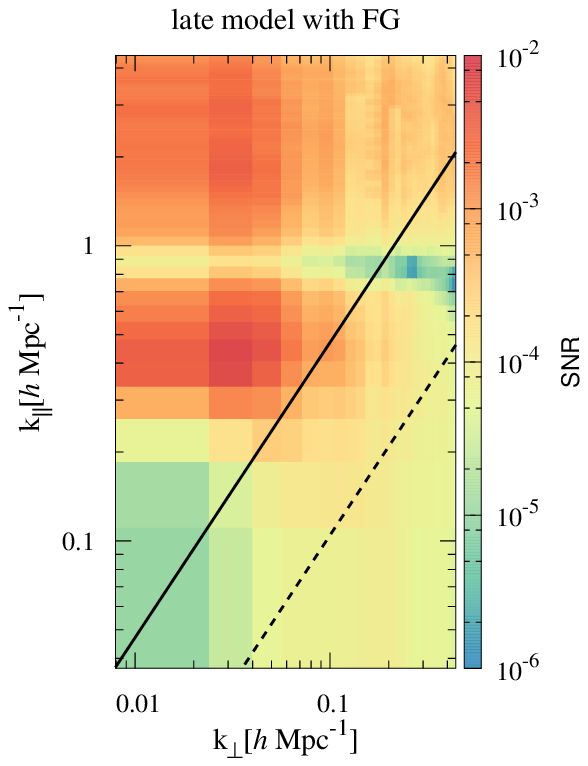}
\caption{{The SNR including foreground contamination for the mid and late models. As anticipated, the SNR is small in the foreground wedge.}} 
\label{fig:SNR} 
\end{figure*}

Including the foregrounds drastically changes the structure of SNR as shown in Fig.~\ref{fig:SNR}. A decrease in the SNR at all scales indicates the foreground PS term dominates the error budget. Especially in the wedge, where the SNR is clearly reduced by the foregrounds and becomes $<10^{-3}$. In the EoR window, foreground contamination is far less, and the SNR remains at $\sim0.01$.

For the mid model, the turn over scale is $k\sim0.3\,h\,\rm Mpc^{-1}$ which is in foreground wedge. Therefore, highly precise foreground removal is required to identify the turn over scale. For the late model on the other hand, the turn over scale is in EoR window, and therefore there is a chance to measure the scale. {The turn over at $k_{\perp}=0.8\,h\,\rm Mpc^{-1}$ is out of plot and lies within the FG wedge. The detection of high $k_{\perp}$ modes requires an array layout with a dense long baseline distribution, which is difficult in present and planned telescopes.} 

\subsection{1D Power Spectrum}\label{sec:1Dpk}
In order to reduce the error, we consider the spherically averaged PS. To convert from a 2D PS into a 1D PS, we take an average of the signal in same $|{\bold{k}}|=k$ bins and use Eq.~\ref{2D21D} for the error estimation. We exclude regions in the foreground wedge from the calculation to minimize contamination by the foregrounds. 

\begin{figure}
\centering
\includegraphics[width=6.0cm]{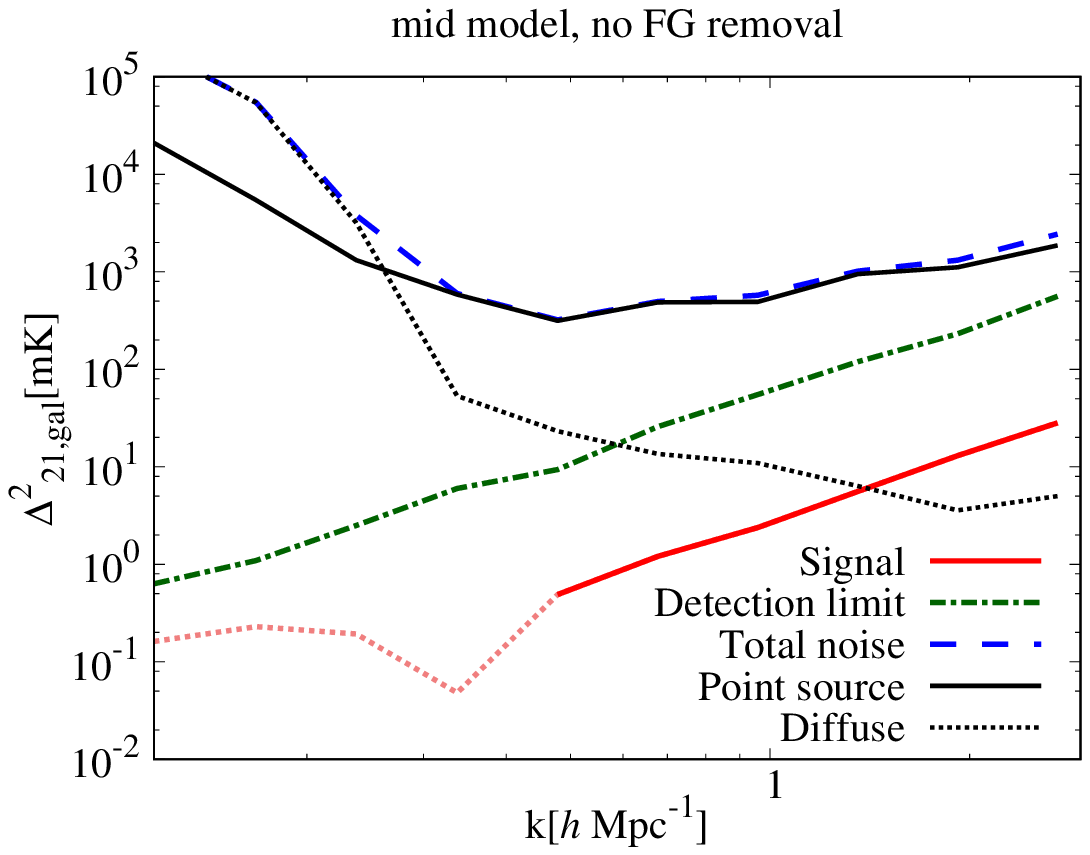}
\includegraphics[width=6.0cm]{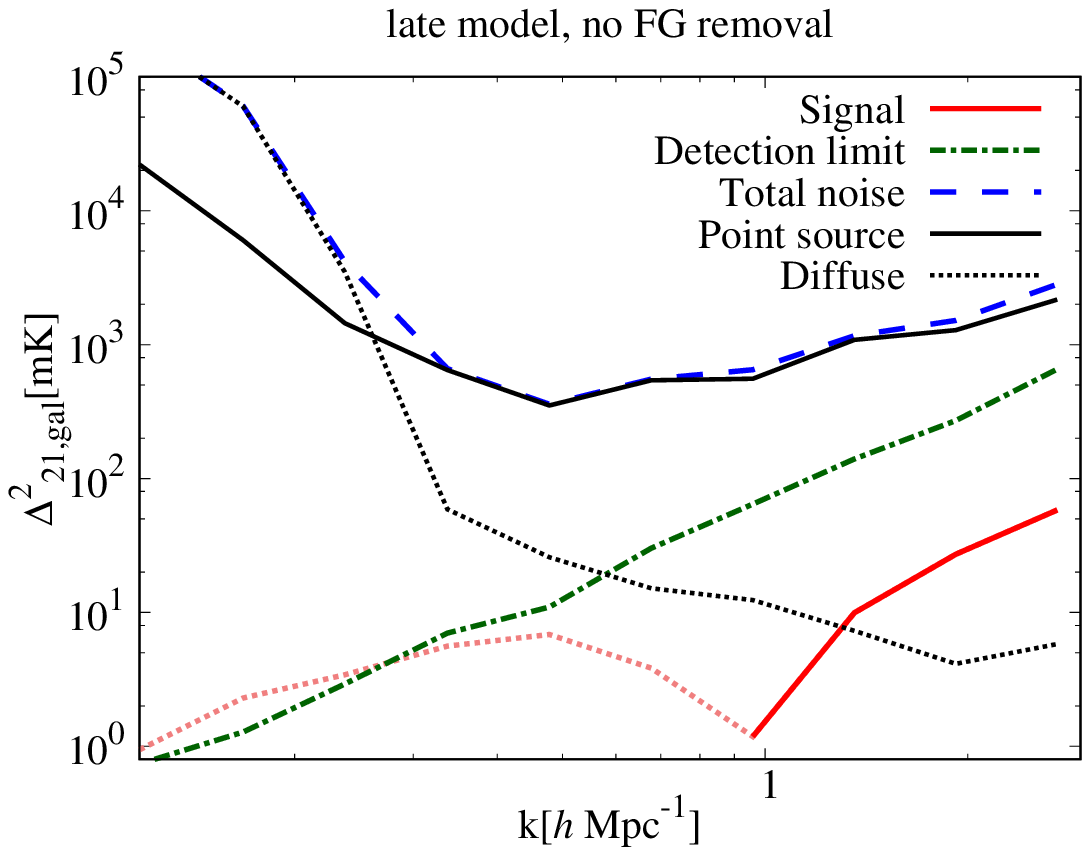} 
\caption{{1D PS calculated from the 2D PS. The mid model is on top panel, late is on bottom. The red line shows the 21$\,$cm-LAE cross PS and negative part is the dashed line, positive is solid line. The dot-dashed line is the detection limit and the dashed line is total error. The solid line shows the contribution from point sources, and the dotted from diffuse emission.}}
\label{fig:1Dcross} 
\end{figure}

The plots in Fig.~\ref{fig:1Dcross} compare the signal and error of 1D PS, where the detection limit represents the term of the thermal noise and the shot noise, which is inevitable error on the observations. For the late model, the signal is larger than the detection limit and therefore the cross PS can be observed by the MWA at large scales ($k\sim0.1\,h\,\rm Mpc^{-1}$) if the foregrounds are perfectly removed as expected in Fig.~\ref{fig:nofgSNR}. However, the total error including the foreground term overwhelms the signal at all scales. The solid and dotted lines show contributions from point sources and diffuse foreground. As we can see, the total error is dominated by diffuse foregrounds at $k\sim0.1\,h\,\rm Mpc^{-1}$ and by point sources at $0.3<k<3.0\,h\,\rm Mpc^{-1}$. Thus, for measuring the signal at large scales, we need to remove the foregrounds by a few orders of magnitudes. Conversely to the large scales, the contribution from thermal noise becomes important at small scale because the foregrounds are reduced at $k_{\parallel}\sim1$ in the EoR window. However, although the decrease of foreground in the EoR window, the contamination from point sources is still 2 order of magnitude larger than the signal. 

\begin{figure}
\centering
\includegraphics[width=6.0cm]{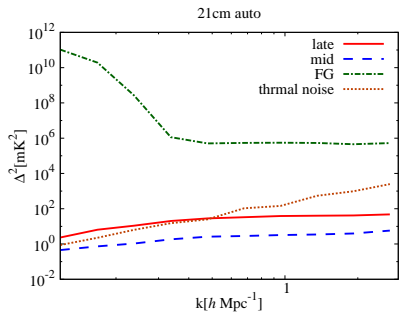}
\caption{{The 1D 21$\,$cm auto PS of the mid and late models are shown as solid and dashed lines, respectively. The dotted line is the thermal noise of MWA 128 tiles. The dot-dashed line is the PS of foregrounds, including point sources and diffuse emission. These are calculated from the 2D PS at $k_{\perp}<0.04$, excluding signals in foreground wedge.}} 
\label{fig:1D21cm} 
\end{figure}

The auto PS is a primary observable of the 21$\,$cm signal. In Fig~\ref{fig:1D21cm}, we compare the 21$\,$cm auto PS for the mid and late models, with the expected thermal noise and foreground PS. {Although the foregrounds are reduced by spherical averaging in the 2D plane for the 21$\,$cm-LAE cross PS, the foregrounds are correlated and do not decrease with averaging for the 21$\,$cm auto PS. }
Thus, we only use the signal in the EoR window at $k_{\perp}<0.08~h\,\rm Mpc^{-1}$, where the foregrounds are naturally avoided. Even if we use such a foreground quiet region, the foreground PS is 4 orders of magnitude larger than the 21$\,$cm signal at $k>0.4~h\,\rm Mpc^{-1}$.

\begin{figure}
\centering
\includegraphics[width=6.0cm]{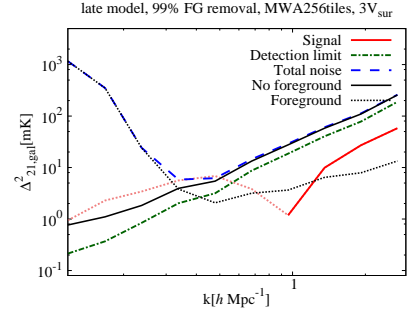}
\caption{{The 1D cross PS of the signal compared with the error, where 99\% foreground removal, the MWA having 256 tiles, and a 3 times larger LAE survey area are assumed. As in Fig.~\ref{fig:1Dcross}, the red line shows the 21$\,$cm-LAE cross PS. The dot-dashed line is the detection limit and the dashed line is the total error. The dotted line is the contribution from foregrounds, which is the dominant term of total error at large scales. The solid line represents the error with prefect foreground removal. The main component of the solid line is thermal noise, which dominates the error at small scales. }}
\label{fig:1Drmhex} 
\end{figure}

\subsection{Requirement for detection}\label{sec:require}
As shown in this section, our results indicate that the foreground removal of a few orders of magnitude is required to detect the 21$\,$cm signal, even if we combine foreground avoidance and cross correlation techniques. In the rest of this section, we discuss how much of the foregrounds we need to subtract. Here, we mainly focus on the late model to find the minimum effort required for measuring the 21$\,$cm signal. 

We have remarked that the error of cross PS is determined by the product of contributions from 21$\,$cm - LAE observations. We have assumed the upcoming LAE survey with the HSC-PFS experiment, which already has the lowest conceivable noise to come in experiments scheduled for the next few years. The only way then to improve the shot noise is to increase the survey area. We therefore mainly discuss the improvement of the thermal noise and foreground removal in the following. It should also be noted that the redshift error of the HSC survey corresponds to the survey depth, $\Delta z\sim0.1$. This large redshift error increases the shot noise; \cite{Kubota} shows that a signal detection at small scales is difficult without the PFS, even using SKA\_{}LOW.

As shown in Fig.~\ref{fig:1Dcross}, the error term including the foregrounds is at least two orders of magnitude larger than signal. At $k>0.4\,h\,\rm Mpc^{-1}$, the contribution from point sources dominates and therefore we need to subtract 99\% of the point source foreground. Although the contribution from diffuse emission is weaker than that from point sources at these scales, we need to remove 80\% of diffuse emission at $k\sim0.4\,h\,\rm Mpc^{-1}$. The level of foreground removal of diffuse emission seem to be possible. For example, in \cite{2016ApJ...833..102B}, they succeeded in removing around 70\% of the diffuse emission. {To achieve these levels they created a diffuse foreground image confined to the main beam of the MWA (of order 20$^\circ$ across) from 3$\,$hours of data. This image was also integrated in frequency, and then subtracted from the data. Their method is relatively simple, and by including spectral structure, and a bright galactic model as well, the foreground removal can be improved. Furthermore, }once they subtracted a point source model based on a hybrid sky catalogue, the power in the wedge was reduced to 2 orders of magnitude weaker than that shown by our point source model in Fig.~\ref{fig:Jpfg}. This indicates 90\% of the point source foreground was removed. Therefore, the precision of foreground removal of point sources which has been already achieved is one order of magnitude worse than to the required level in this work.

While we find hopeful results at $k<0.3~\,h\,\rm Mpc^{-1}$, there is serious foreground leakage into the EoR window, and the SNR is less than $10^{-4}$ at $k\sim0.2\,h\,\rm Mpc^{-1}$. There, we need a 99.99\% reduction of the diffuse emission. Although the error can be reduced if we can increase the survey volume of LAE observations, we need to increase the survey area by a factor of 100 to achieve a 90\% reduction of the total error. Detection of the cross PS then necessitates exquisite foreground removal because such an extremely large survey area is not realistic. 

As in the case of the cross PS, the detection of the 21$\,$cm auto PS requires foreground removal. The foreground PS is 4 orders of magnitude larger than the signal at $k>0.3\,h\,\rm Mpc^{-1}$ as we can see in Fig.~\ref{fig:1D21cm}. Note that the unit of auto PS is the square of mK. Thus, 99\% of the foregrounds have to be removed, although the avoidance technique effectively reduces the foreground. The required precision for foreground removal is equal to or higher than that of the case of cross PS. 

We need not only foreground removal but also high sensitivity at small scales for measuring the signal. Especially at $k\sim2\,h\,\rm Mpc^{-1}$ where the error is strongly contaminated by thermal noise, which is 9 times larger than the signal. To reduce the thermal noise error of the cross PS by a factor of 9, 81000 hours of integration time with 128 MWA tiles, or 256 tiles for 20000 hours, are required. Future telescopes, for example the MWA phase 2, HERA, and SKA\_{}LOW, have higher sensitivities and may be able to detect the cross PS at all scales. 

In Fig.~\ref{fig:1Drmhex}, we demonstrate the minimum requirement for detection of cross PS. {We mention that we do not apply any realistic foreground removal, which is outside the scope of this paper. Here, the foreground PS is reduced by constant factor at all scales.} The foreground PS has been reduced by a factor of 10000, which corresponds to 99\% foreground removal. The thermal noise PS is also reduced by a {constant} factor of {4}, which corresponds to an assumption of using 256 MWA tiles. In addition, we increase the survey volume of the LAE survey by a factor of 3. With these changes, the SNR becomes larger than 1 at $k\sim0.4\,h\,\rm Mpc^{-1}$. Furthermore, the turn over scale can be identified with future telescopes, if 99\% foreground removal and 80 times higher sensitivity of MWA can be achieved. The turn over scale indicates the typical size of ionized bubbles, allowing us to constrain reionization models.

We need to mention some difficulties which have not been taken into account in this work. First, we have simulated observations with the MWA with a smooth bandpass but, in practice, the MWA has a coarse band structure that causes data loss. When the Fourier transform is performed with respect to frequency, coarse band harmonics cause strong foreground contamination in EoR window as horizontal lines \citep[e.g.][]{2016ApJ...818..139T, 2016ApJ...833..102B}. The MWA therefore cannot observe some bins in the 2D PS. {Additionally, it should be noted that the foreground contamination in the EoR window depends on the channel width, bandwidth, and bandpass taper.}  

Next, we assumed perfect calibration and no instrumental error. In practice, there are many systematic difficulties such as gain and phase calibration of the visibilities, imperfect beam model, and cable reflections \cite[e.g.][]{2012ApJ...752..137M,2016MNRAS.461.3135B,2016ApJ...825..114J}. As these errors cause more foreground contamination in the EoR window, higher levels of foreground removal will be required. In order to estimate the effect of these instrumental problems, an end-to-end simulation is required, which we will focus on in future works.

Imperfect point source subtraction leaves residuals and it makes foreground removal less effective~\citep[]{2012ApJ...757..101T,2017PASA...34....3L,2017arXiv170702288P}. The ionosphere refracts the radio waves, which to first order changes the observed position of point sources. Thus, to subtract point sources precisely, we need to calibrate the ionosphere every few seconds.  

\section{Summary}\label{sec:summary}
In this work, we have studied the effects of foregrounds on the observation of 21$\,$cm-LAE cross PS. Although the LAEs and foregrounds have no correlation, the PS of foregrounds contribute to the error of 21$\,$cm-LAE PS. To estimate the PS of the foregrounds, we have assumed a realistic MWA observation, including the pointing-dependent beam shape, and a model of extra galactic point sources consisting of radio galaxies found in the GLEAM catalogue. We also use a parametric model of diffuse emission from our Galaxy. The 21$\,$cm signal is calculated using a numerical hybrid simulation combining an RHD simulation with post processed radiative transfer. We identified LAEs using a Lyman-$\alpha$ transmission code. We also assume a LAE survey with HSC and the PFS spectroscopic observations. 

We found that the foreground contribution to the error is larger than the thermal noise contribution at large spatial scales, although the foreground is reduced in the EoR window. This foreground contamination is inevitable even if we use future telescopes such as MWA phase 2, HERA and SKA\_{}LOW. We therefore must remove the foregrounds to detect the cross PS. However, the required precision of foreground removal is $\sim99$\% of point sources and only 80\% of diffuse emission at $k \sim 0.4 h\,\rm Mpc^{-1}$. As with the large scales, the 99\% of point sources removal is necessary to detect the signal at small scales $k>1h\,\rm Mpc^{-1}$. We mention that the contamination in the EoR window can be reduced by the dense $uv$-coverage and therefore the required precision can be mild for future telescopes. The error term of thermal noise and shot noise is larger than the expected signal at especially small scales and therefore we need a sensitivity of at least 80 times greater than that of MWA with 128 tiles. 

{
We mention three crucial problems which are out of the scope of this paper. 
First, in this work, the specific method of foreground removal was not studied. 
The measurement of the signal needs effective methods which can remove foregrounds to the levels required.
In addition, we focused on the MWA but different instruments such as HERA and SKA\_{}LOW change the feasibility of reaching the required foreground removal limits. 
Finally, the instrumental and calibration errors pointed out in section \ref{sec:require} can be fatal obstacles.
These problems will be studied in our future works. 
}

%

\section*{Acknowledgement}
We would like to thank Cathryn Trott for helpful comments that improved the paper, Rachel Webster and Bart Pindor for useful discussion of our first article, Masami Ouchi for guidance of HSC-PFS survey, Hidenobu Yajima for  providing us with the Lyman-$\alpha$ transmission code and Tomoaki Ishiyama for conducting the $N$-body simulation used in this work. This work was supported by resources awarded under Astronomy Australia Ltds merit allocation scheme on the gSTAR national facility at Swinburne University of Technology. gSTAR is funded by Swinburne and the Australian Governments Education Investment Fund. This work is supported by Grant-in-Aid from the Ministry of Education, Culture, Sports, Science and Technology (MEXT) of Japan, No. 16J01585 (SY)., No. 26610048 (KT),  No.15H05896 (KT),  No.16H05999 (KT), No.17H01110 (KT), and Bilateral Joint Research Projects of JSPS (KT).

\end{document}